\documentclass[11pt,twoside,a4paper]{article}
\usepackage[a4paper, total={6in, 9in}]{geometry}

\usepackage[utf8]{inputenc}
\usepackage[english]{babel}
\usepackage[T1]{fontenc}
\usepackage{amsmath}
\usepackage{amsthm}
\usepackage{amsfonts}
\usepackage{hyperref}
\usepackage{graphicx}
\usepackage{braket}
\usepackage{enumerate}
\usepackage{multirow}
\usepackage{tikz}
\usepackage{algorithm}
\usepackage{algorithmic}
\usetikzlibrary{quantikz}

\usepackage{caption}
\usepackage{subcaption}

\usepackage{algorithm,algorithmic}
\usepackage{balance}

\newtheorem{theorem}{Theorem}
\newtheorem{lemma}[theorem]{Lemma}

\begin{document}

\begin{center}
{\Large
\textbf\newline\newline\newline{Resource-aware scheduling of multiple quantum circuits\\ on a hardware device}
}
\end{center}

\begin{center}
Debasmita Bhoumik \textsuperscript{1,*},
Ritajit Majumdar \textsuperscript{2},\\
Susmita Sur-Kolay\textsuperscript{1}
\\
\bigskip
\textsuperscript{1} Advanced Computing \& Microelectronics Unit,\\ Indian Statistical Institute, India\\
\textsuperscript{2} \emph{IBM Quantum}, IBM India Research Lab\\
* debasmita.ria21@gmail.com

\end{center}

\begin{abstract}
Recent quantum technologies and quantum error-correcting codes emphasize the requirement for arranging interacting qubits in a nearest-neighbor (NN) configuration  while mapping  a quantum circuit onto a given hardware device, in order to avoid undesirable noise. It is equally important to minimize the wastage of qubits in a quantum hardware device with $m$ qubits while running circuits of $n$ qubits in total, with $n < m $. In order to prevent cross-talk between two circuits, a buffer distance between their layouts is needed. Furthermore, not all the qubits and all the two-qubit interactions are at the same noise-level. Scheduling multiple circuits on the same hardware may create a possibility that some circuits are executed on a noisier layout than the others.

In this paper, we consider an optimization problem which schedules as many circuits as possible for execution in parallel on the hardware, while maintaining a pre-defined layout quality for each. An integer linear programming formulation  to ensure maximum fidelity while preserving the nearest neighbor arrangement among interacting qubits is presented. Our assertion is supported by comprehensive investigations involving various well-known quantum circuit benchmarks. As this scheduling problem is shown to be NP Hard, we also propose a greedy heuristic method which provides 2$\times $ and 3$\times $ better utilization for 27-qubit and 127-qubit hardware devices respectively in terms of qubits and time. 

\end{abstract}



\section{ Introduction}
In the past decade, quantum computation has matured from a mere theoretical model of computation which can outperform traditional computers on certain problems of interest \cite{shor1994algorithms, Grover:1996:FQM:237814.237866} to performing experiments on a real quantum computer beyond brute force classical computation \cite{kim2023evidence}. Currently multiple providers offer free or paid access to their quantum hardware devices. The majority of these access are via cloud. 
As the demand for quantum computing is rising, the users often face long queuing time with their jobs having to wait till the execution of all the previously submitted jobs have been completed. Consequently, there is a pressing need to enhance the efficiency and throughput of quantum computers to improve user experience. In this paper, we study the benefits and challenges of executing more than one quantum circuit simultaneously on the same hardware to improve the throughput, as well as the hardware utilization, without sacrificing on the quality of outcome.

\begin{figure}[H]
    \centering
    \includegraphics[scale = 0.2] {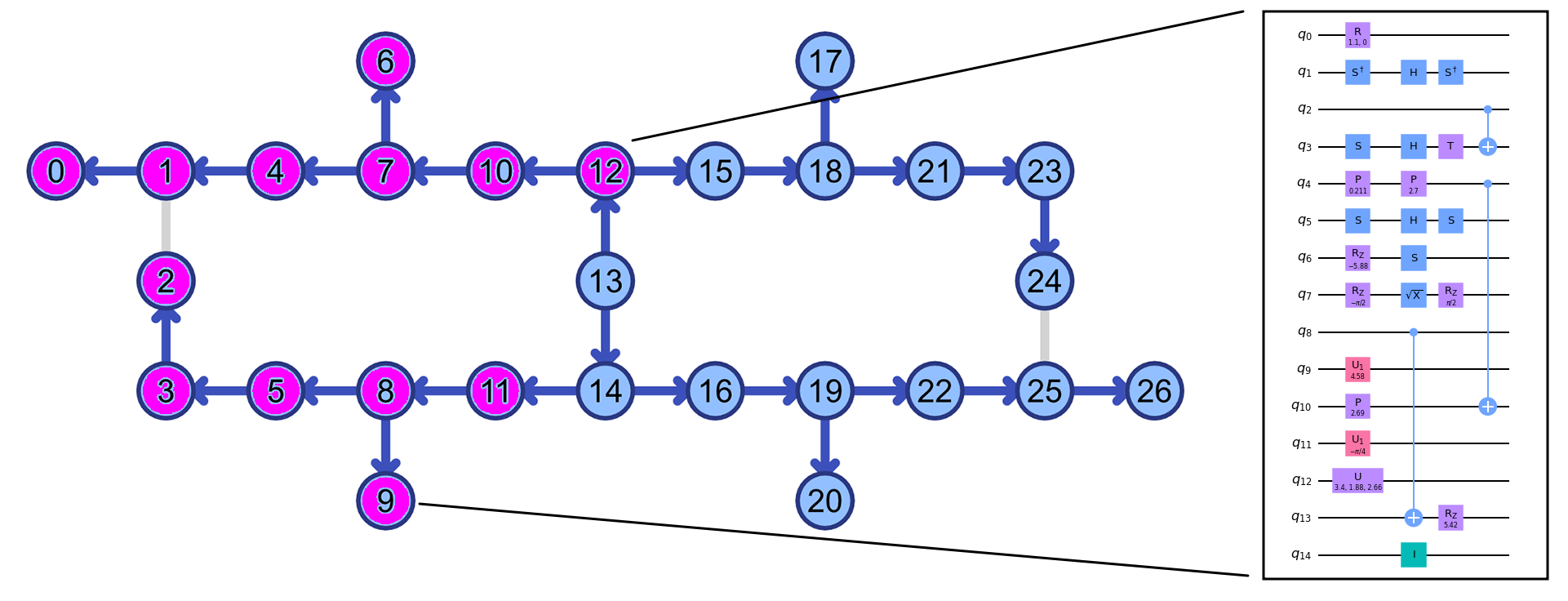}
    \caption{An example of a 15-qubit circuit assigned to a 27- qubit hardware. The used qubits are shown in purple while the unused qubits are shown in blue. The hardware still has room to accommodate one or more quantum circuit(s) using the free qubits.}
    \label{fig:lessutil}
\end{figure}

Let us motivate the problem with an example. Consider a 15-qubit circuit which is to be executed on a 27 qubit device as shown in Fig~\ref{fig:lessutil}. Thus 12 qubits of the device remain unused, which could have been utilized to  execute simultaneously some other circuit(s) requiring $\leq 12$ qubits -- thus improving the throughput and the hardware utilization.

Simultaneous execution of multiple circuits on a single hardware is not without challenge. Previous studies \cite{cross2017open} do not consider the effect of noise arising due to simultaneous execution of circuits. First, when a circuit is mapped to a hardware, the requirement is to reduce the number of SWAP gates, as well as to use a layout with minimal noise profile \cite{treinish2022mapomatic}. However, when multiple circuits are placed simultaneously, it is   likely that all of them cannot be placed on their corresponding best layout, leading to degradation in the quality of the outcome of the computation. Furthermore, if two  circuits are computed on neighbouring qubits, then there is a possibility of crosstalk affecting the quality of the computation for both of them.

Minimizing the degradation in quality due to worse layout selection and crosstalk, while maximizing the throughput of the hardware lead to conflicting objectives. In this paper, we study this optimization problem to find the optimal \emph{intra-device scheduling} of $N > 1$ jobs on a $m$-qubit hardware with little to no compromise on the quality of computation. Given a hardware $H$, and a set $C$ of $N$ circuits $\{C_1, C_2, ..., C_N\}$, such that the number of qubits of each circuit is $\leq$ the number of qubits in the hardware, we partition $C$ into batches $B = \{B_1, B_2, ..., B_k\}$ such that each batch consists of circuits which can be executed simultaneously. Therefore, for each $i$, we must have $1 \leq |B_i| \leq N$ where $|B_i|$ indicates the number of circuits in that batch and the total number of qubits in each $B_i$ is no greater than $m$. Maximizing intra-device parallelization is thus equivalent to minimizing the number of batches.
Our contributions in this work are as follows:

\begin{itemize}
    \item  obtaining an optimal placement of more than one circuit on a given hardware device using  integer linear program (ILP) such that the noise profile of the layout of each circuit is within a pre-specified margin of  $\epsilon \ll 1$ in the noise profile of the optimal layout, and a buffer distance $b$ is maintained between any two distinct circuits executed in parallel to avoid crosstalk;
    \item given a set of circuits, finding optimal schedule batches of circuits with mapping on the hardware a minimum  number of batches;
    \item proving that this optimal scheduling problem is NP-Hard, and hence designing a heuristic graph-theoretic approach to solve this problem;
    \item presenting experimental results on 27-qubit fake backend and 127-qubit quantum hardware in which $2\times$ and $3\times$ increase in the throughput are obtained respectively with only a reduction in mean fidelity by $\sim 1.4\%$ for 10-qubit circuits.
\end{itemize}

In the rest of the paper, Sec.~\ref{sec:background} provides a brief discussion on circuit placement and the noise profile of a layout on a given hardware device. Sec.~\ref{sec:framework} delves in the proposed methodology for intra-device-scheduling. Our experimental settings and results appear in Sec.~\ref{sec:results} and concluding remarks in Sec.~\ref{sec:conclusion}.

\section{A brief introduction to circuit mapping and layout scoring}
\label{sec:background}

\begin{figure} [ht]
    \centering
    
    \includegraphics[scale = 0.5] {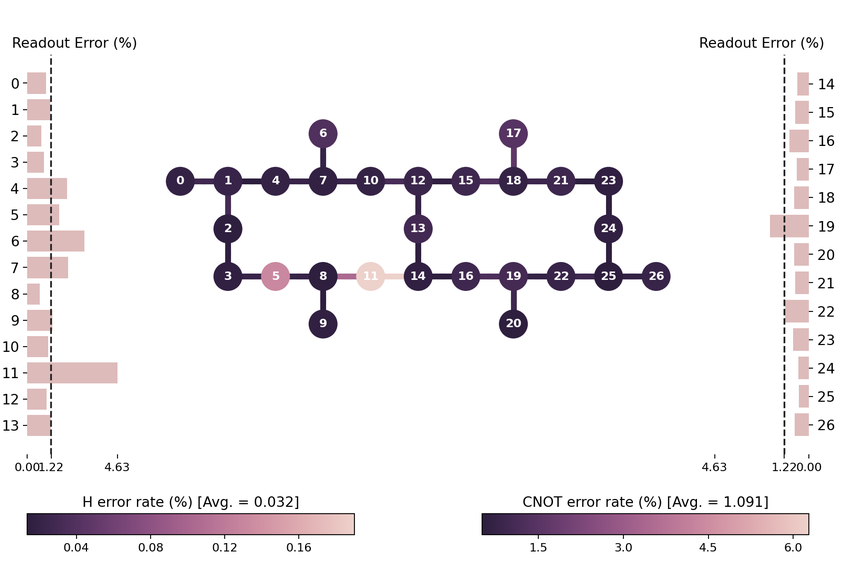}
    \caption{The coupling map and bit-wise noise profile of a 27-qubit IBM Quantum hardware}
    \label{fig:ibmkolkata}
\end{figure}

The virtual qubits of a quantum circuit need to be mapped to the physical qubits of the hardware for execution. Current superconductor based quantum hardware typically have a planar architecture, leading to a sparse coupling map. Fig.~\ref{fig:ibmkolkata} depicts the coupling map of a 27-qubit IBM Quantum device, which is a sparse graph with degree-2 and degree-3 connectivity.

Two qubit interaction is possible only between neighbouring qubits. If any interaction is required between non-neighbouring qubits, then SWAP gates are required to make them adjacent. Adding these SWAP gates increases the computation time as well as the noise in the circuit. Therefore, the primary constraint of mapping the virtual qubits to the physical qubits on a given hardware is to minimize the number of SWAP gates. This being an NP-Hard problem, multiple heuristics have been proposed \cite{li2019tackling, sivarajah2020t, kremer2024practical}. In this study, we employ the default SABRE algorithm \cite{li2019tackling} used in Qiskit \cite{Qiskit}.

A second constraint on qubit mapping is to use qubits and 2-qubit connections which have lower noise profiles. For example, the color code in Fig.~\ref{fig:ibmkolkata} shows that qubit 11 is significantly noisy, and it is advantageous to exclude this qubit, if possible, during mapping. Certain studies consider the minimization of SWAP gates and noise profile as a single objective function \cite{murali2019noise, wille2019mapping}. In \cite{treinish2022mapomatic}, the authors proposed a two-step solution for the same. First, they map the virtual qubits to the physical qubits of the hardware without considering the noise profile of the qubits and the connections. After obtaining a layout which minimizes the number of SWAP gates, the authors find a list of isomorphic layouts which keep the number of SWAP gates unchanged. Finally, the noise profile of each layout is calculated from the noise calibration data, and the one with the lowest noise profile is selected. This entire process is termed as{\it mapomatic} by the authors.

Mapomatic is an open-source package in \href{https://github.com/qiskit-community/mapomatic/tree/main}{Qiskit Community}. Given a circuit $C_i$, we use this method to obtain the optimal qubit placement, and a list $L$ of isomorphic layouts  with $Q_{ij}$ denoting the score of layout $l_j \in L$. In other words, $Q_{ij}$ is an indicator of the noise profile of the layout $l_j$ for the circuit $C_i$. In general, the scoring mechanism of mapomatic ensures that $0 \leq Q_{ij} \leq 1$, and if $Q_{ij} < Q_{ij'}$, then $l_j$ is considered to be a better layout than $l_{j'}$ in terms of noise profile.

\section{Formulation of intra-device scheduling as an optimization problem}
\label{sec:framework}
Consider a hardware $H$ with $m$ qubits and a set $C = \{C_1, C_2, ..., C_N\}$ of quantum circuits. The goal of this study is to place batches of $k \leq N$ circuits simultaneously on the hardware such that (i) the total number of qubits for each batch of circuits is no greater than $m$, (ii) there is no overlap of qubits between different circuits, and (iii) the noise profile of the layout associated with each circuit is within $\epsilon$ of the optimal layout for that circuit. In other words, if $s_1$ be the noise profile of the optimal layout for a circuit, then for that circuit we  consider only those layouts $l$ for which $s_l - s_1 < \epsilon$, for a pre-specified $\epsilon$.

\begin{figure} [ht]
\centering
\includegraphics[scale = 0.18] {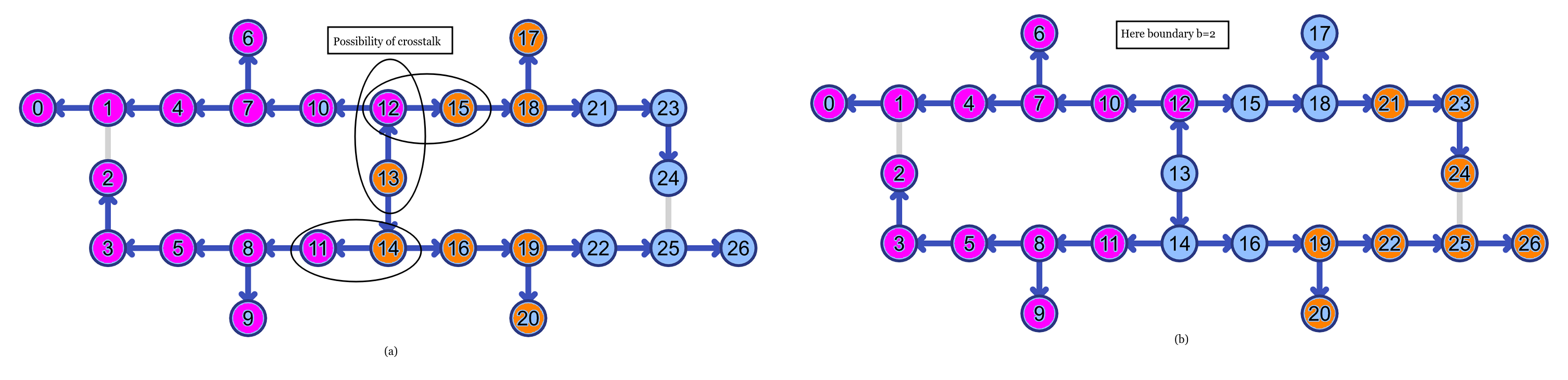}
\caption{Placement of a 15-qubit and a 8-qubit circuit simultaneously on a 27-qubit hardware with buffer distance (a) $b = 0$, and (b) $b = 2$. The former has a significantly higher probability of crosstalk affecting the quality of the computation. The blue qubits are the unused ones.}
\label{fig:boundary}
\end{figure}

Note that the  constraint of no overlap between qubits may seem trivial at first. However, it is to be noted that if two neighbouring qubits are associated with different circuits, then there is a possibility of crosstalk affecting the quality of outcome of both the circuits. Therefore, a buffer distance $b$, i.e., a distance of $b$, must be maintained between any two qubits associated with two different circuits. In Fig.~\ref{fig:boundary}, we show examples of two circuits placed with $b=0$ and $b=2$ respectively. The former has a significantly higher possibility of crosstalk affecting the quality of outcome \cite{murali2019noise}.

Let $l_1$ and $l_2$ be two layouts for two circuits $C_1$ and $C_2$. Let the distance between two qubits $q_a \in l_1$ and $q_b \in l_2$ on the same hardware be denoted by $d(q_a, q_b)$. Henceforth, for a given buffer distance $b$ we say that $l_1$ and $l_2$ have $b$-overlap  if $\exists q_a \in l_1$ and $\exists q_b \in l_2$ such that $d(q_a, q_b) < b$,  else if $\forall$ $q_a \in l_1$ and $q_b \in l_2$, $\min d(q_a, q_b) \ge b$, then $l_1$ and $l_2$ are $b$-non-overlapping. 
In the next subsection, we propose an efficient algorithm to determine the overlap between  two given layouts.

\subsection{Finding the overlap between two layouts}
Given a hardware coupling map, and two layouts $l_1$ and $l_2$, the trivial method to check for overlaps is to calculate $d(q_a, q_b)$, $\forall$ $q_a \in l_1$ and $q_b \in l_2$. However, if the  number of qubits in $l_1$ and $l_2$ are $n_1$ and $n_2$ respectively, then the time complexity of this method is $\mathcal{O}(n_1.n_2)$. But checking for overlaps only between \emph{boundary qubits} of the two layouts can be done faster.

A qubit in a layout $l$ is said to be a \emph{boundary qubit} if it is adjacent to at least one qubit $q \notin l$. For example, in Fig.~\ref{fig:boundary} (b), qubits 11, 12, 21 and 19 are the boundary qubits.

\begin{lemma}
\label{thm:overlap}
    The overlap between the two layouts $l_1$ and $l_2$ can be determined by calculating the distance between the boundary qubits of these two layouts only.
\end{lemma}

\emph{Proof}: See Appendix~\ref{app:overlap}.

In Fig.~\ref{fig:boundary} (b), there are 13 and 8 qubits on the two layouts. Therefore, the naive method would have required $13 \cdot 8 = 104$ comparisons to determine the overlap. However, since there are only 2 boundary qubits for each layout, according to Theorem~\ref{thm:overlap} only 4 comparisons are sufficient.

Next we provide the algorithms for determining the boundary qubits of a layout, and calculating the overlap between two given layouts.

\begin{algorithm}
\caption{Determine boundary qubits of a layout}
\label{alg:find_boundary}
\begin{algorithmic}[1]
\REQUIRE Layout $L_i$ and the coupling map of the hardware
\ENSURE A list of the boundary qubits of $L_i$
\STATE $boundary \leftarrow []$
\FOR{each qubit $q \in L_i$}
    \STATE Determine the neighbours $Nb$ of $q$ from the coupling map
    \FOR{each qubit $q' \in Nb$}
        \IF{$q' \notin L_i$}
            \STATE Add $q'$ to $boundary$
        \ENDIF
    \ENDFOR
\ENDFOR
\RETURN $boundary$
\end{algorithmic}
\end{algorithm}

\begin{lemma}
\label{thm:find_boundary}
Algorithm~\ref{alg:find_boundary} finds the boundary qubits of a layout $l$ with $n_l$ qubits in $\mathcal{O}(n_l)$.
\end{lemma}

\emph{Proof}: See Appendix~\ref{app:find_boundary}.

\begin{algorithm}
\caption{Check $b$-overlap between two layouts}
\label{alg:check_overlap}
\begin{algorithmic}[1]
\REQUIRE Layouts $l_i$, $l_j$, coupling map $M$ of the hardware, desired buffer distance $b$
\ENSURE True if $overlap$, False otherwise
\STATE $overlap \leftarrow$ False
\IF{ $l_i \cap_0 l_j \neq \phi$ }
    \STATE $overlap \leftarrow$ True
    \RETURN $overlap$
\ENDIF
\STATE $boundary_1 \leftarrow$ find\_boundary($l_i$, $M$)
\STATE $boundary_2 \leftarrow$ find\_boundary($l_j$, $M$)

\FOR{each qubit $q_i$ in $boundary_1$}
    \FOR{each qubit $q_j$ in $boundary_2$}
        \STATE $d  \leftarrow$  distance($q_i$, $q_j$)
        \IF{$d  \leq b$}
            \STATE $overlap \leftarrow$ True
            \STATE \textbf{break}
        \ENDIF
    \ENDFOR
\ENDFOR

\RETURN $overlap$
\end{algorithmic}
\end{algorithm}

\begin{lemma}
\label{thm:check_overlap}
    Given two layouts $l_i$ and $l_j$, with $n_i$ and $n_j$ qubits of which $k_i$ and $k_j$ denote respectively the number of boundary qubits of the two layouts, then Algorithm~\ref{alg:check_overlap} finds the overlap between the two layouts in time $\mathcal{O}(\max \{n_i, n_j\} + k_i \cdot k_j)$.
\end{lemma}

\emph{Proof}: See Appendix~\ref{app:check_overlap}.

Given a circuit $i$ and a hardware $H$, let $L$ be the list of all isomorphic layouts obtained from mapomatic. Let $l \in L$ be the best layout with the lowest mapomatic score. We define a subset $L_{\epsilon} \subseteq L$ such that for each $l' \in L_{\epsilon}$, $Q_{il'}-Q_{il} \leq \epsilon$. Recall that by definition, the lower the mapomatic score, the better is the layout. 

The problem at hand, therefore, is to schedule a set of circuits $C$ to a set of layouts $L_{\epsilon}$ for each circuit such that no two layouts corresponding to two different circuits are $b$-overlapping. The theoretical formulation and the solution of this problem does not depend on a specific value of $\epsilon$. Therefore, for the rest of the paper, we  exclude any explicit mention of $\epsilon$. However, whenever a list of layouts $L$ is mentioned, it  implies $L_{\epsilon}$ for a pre-specified $\epsilon$. In Sec.~\ref{sec:results} we shall discuss the choice of $\epsilon$ for our experimental results.

\subsection{ILP Formulation for our scheduling problem}

Let us formulate an integer linear program (ILP) for the noise-aware intra device scheduling (NIDS). Consider a list of circuits $C = \{C_1, C_2, ..., C_N\}$, and a list of hardware layouts $L_{all} = L_1 \cup L_2 \cup ... \cup L_N$, where $L_i$ denote the layouts feasible for circuit $C_i$. For example, in Fig.~\ref{fig:boundary} (b), the two layouts differ in the number of qubits, therefore the layout for one circuit is not feasible for the other. However, there may also be cases, where the two circuits $i$ and $j$ have equal number of qubits, and therefore the same layout may belong to both $L_i$ and $L_j$. For the rest of this paper, we shall remove the index, and simply use $L$ to denote the list of layouts for any circuit $i$. The index is same as that for  the circuit in the context.

Furthermore, it may not be possible to accommodate all the circuits simultaneously on the hardware. This may be because there are not sufficient quality layouts, or the total number of qubits required exceeds the number of physical qubits in the hardware. Therefore, the goal is to accommodate the largest subset of circuits simultaneously, and repeat this process of intra-device scheduling to schedule all the circuits in $k$ batches $B_1, B_2, ..., B_k$, where each batch is a set of circuits which can be executed simultaneously, and $k$ is expected to be significantly smaller than $N$. Note that this method can accommodate for instances where the circuits are added to the job queue dynamically.

Next we discuss the variables, constraints and the objective function of the ILP for scheduling the largest subset of circuits simultaneously on a hardware.

\begin{enumerate}
    \item \textbf{Variables}
    \begin{enumerate}
        \item \emph{Indicator variables}: We associate $x_{ij}$ for each circuit $i \in C$ and layout $j \in L$ such that
$$
x_{ij} = \begin{cases}
    1 & \text{if circuit $C_i$ is scheduled to layout $j$} \\
    0 & \text{otherwise}.
\end{cases}
$$

        \item \emph{Score variables}: A score variable $q_{ij}$ is associated with each $x_{ij}$ which is the mapomatic score when circuit $C_i$ is placed on layout $j$.
    \end{enumerate}

\item \textbf{Constraints}\label{constraint} 

\begin{enumerate}

 \item  Since $x_{ij}$ are indicator variables, we require that $\forall$ $i, j$
    \begin{equation}
     \label{eq:constraint3}
     x_{ij} \in \{0,1\}
     \end{equation}

    \item The second requirement is that every circuit $C_i$ is assigned to at most one layout. Note that if a circuit is not scheduled in a particular batch, then it is not assigned any layout, and hence $\forall$ $j$, $x_{ij} = 0$. Formally, this constraint can be represented as 
        \begin{equation}
         \label{eq:constraint1}
         \sum_{j \in L} x_{ij} \leq 1
        \end{equation}
    Note that this constraint should hold for all circuits $C_i \in C$, so there are $|C|$ such constraints.

    \item 
        The third requirement is that no circuit is placed on a layout which has overlap with a previously mapped circuit, i.e., all circuits must be mapped to non-intersecting layouts.
    \begin{equation}
     \label{eq:constraint2}
     x_{ij} + \sum_{k \neq i \in C} \sum_{j \cap_b l \neq \phi, j,l \in L} x_{kl} \leq 1
     \end{equation}
     This constraint implies that two circuits $i$ and $k$ should not be placed on $b$-overlapping layouts. As before, this constraint should hold for all circuits $i \in C$, and therefore, there are $|C|$ such constraints.

\end{enumerate}

\item \textbf{Objective Function}: The objective of this optimization problem is to maximize the overall fidelity, which translates to minimizing the overall score $Q$ along with maimising resource utilisation on the given hardware device. Therefore, the objective function is given by:
\begin{equation}
     \label{eq:obj}
Minimize \sum_{i \in C,} \sum_{j\in L} q_{ij} A_i x_{ij} - \sum_{i \in C,} \sum_{j\in L} x_{ij}
\end{equation}

 where $A_i$, $0 < A_i \leq 1$  denotes the area of circuit $i$ normalized over all available circuits. The area of a circuit is defined as the product of the number of qubits and its depth. 

Note that the second term ensures that the objective function can have negative values; otherwise the least attainable value would have been 0, which could be attained if no circuit is placed at all.  Inclusion of this term ensures that circuits with higher area (i.e., more amenable to noise) are placed in layouts with lower mapomatic score to ensure quality of outcome. 

\end{enumerate}

The solution to this ILP shall provide the largest subset/batch of circuits $B \subseteq C$ which can be executed simultaneously on a given hardware. When $B \subset C$, the same ILP can be solved for $C \setminus B$ repetitively to schedule all the circuits into batches. Let $B_1, B_2, ..., B_k$ denote the batches, where it is expected that $k < N$. This reduces the number of times the quantum hardware is accessed, and in its turn increases the throughput and the hardware utilization.

Note that in current cloud providers, jobs enter the queue dynamically. Therefore, the set of all circuits $C$ may not be static and known a priori. However, this  can be easily accounted for as follows. At a particular timestamp $T$, let the list of available circuits be $C$,  of which $C'$ has been allocated simultaneously to the hardware. Let $\Bar{C}$ denote $C \setminus C'$. While these circuits in $C'$  are being executed, let $C_T$ be the set of new circuits which are added to the queue. IN the next iteration the ILP is solved on $\Bar{C} \cup C_T$ to find the largest subset of circuits to be sent for simultaneous execution at the next timestep.

\subsubsection{Obtaining the outcomes of the individual circuits}
Let $B = \{c_1, c_2, ..., c_r\}$ be the batch of circuits executed simultaneously on the hardware. If the number of qubits on circuit $c_i$ be $n_i$, then effectively the hardware computes a circuit of $n = \sum_{1 \leq i \leq r} n_i$ qubits. The outcome of this execution will be a probability distribution over $n$ qubits. The outcome of the circuit $c_i$ can now be obtained by marginalizing over the rest of the $n-n_i$ qubits.

For example, in Fig.~\ref{fig:boundary} (b), the hardware effectively executes a circuit with $(13+8) = 21$ qubits. The outcome will be a probability distribution encompassing all the 21 qubits. The distribution of the first circuit (in purple) can now be obtained by marginalizing over the remaining 8 (in orange) qubits, and vice versa. Note that although the hardware executes a single circuit, it is essentially a combination of multiple disjoint circuits with no entanglement flowing from one to the other. Therefore, such a marginalization does not lead to any loss of information.

\subsubsection{Noise-aware intra device scheduling is NP-Hard}
Here we show that intra-device scheduling of $N$ circuits is essentially a bin packing problem. The $N$ circuits are the items. The bins are the batches $B_1, B_2, ..., B_k$ and the capacity of each batch is the total number of qubits on the hardware. Thus the objective of this problem is to minimize the number of batches, each having a capacity equal to the number of qubits on the hardware, to pack $N$ quantum circuits in bins of equal capacity. Since the intra-device scheduling is equivalent to a bin packing problem, it is NP-Hard.

Note that our problem at hand imposes additional constraints to this. It requires that the circuits (i.e., the items) placed in each batch (i.e., each bin) do not have $b$-overlapping layouts. This constraint implies that the layouts of two circuits with $b$-overlap cannot be placed in the same batch even if the capacity supports it. The bin-packing problem is reducible to this problem in polynomial time, therefore our problem is NP-Hard. Further, the problem with the layout buffer constraint is at least as hard as the bin packing problem.

The ILP formulation of this NP-Hard problem being computationally expensive and not scalable, we propose in the next section a polynomial time graph-based  greedy heuristic algorithm for noise-aware intra device scheduling.

\section{Polynomial time heuristic algorithm}
\label{sec:heuristic}

In this section we present a polynomial time heuristic algorithm for solving the noise-aware intra device scheduling problem described in the previous section. First,  we create a compatibility graph for the circuits and their layouts: each vertex denotes a distinct circuit and its corresponding layout, and an edge between two vertices denotes that the two circuits are distinct, with the two layouts not $b$-overlapping. The edges are weighted with a function of the mapomatic score of the two associated vertices. We finally propose a polynomial time heuristic algorithm to determine a \emph{maximal} clique from this compatibility graph, which denotes the set of circuits that can be executed simultaneously. We shall discuss each step of this approach in detail.

\subsection{Generation of the compatibility graph}
The generation of the compatibility graph $G$ is given in Algorithm~\ref{alg:graph_construction_with_weights}. 

\begin{algorithm}[htb]
\caption{Generate compatibility graph}
\label{alg:graph_construction_with_weights}

\begin{algorithmic}[1]
\REQUIRE A set $C$ of circuits ; for each $i \in C$ a list $L_i$ of isomorphic layouts and the normalized circuit area $A_i$; for each $i \in C$ and $j \in L_i$ a mapomatic score $q_{ij}$; a buffer distance $b$
\ENSURE Compatibility graph $G$
\STATE $G \leftarrow$ empty graph.
\FOR{each $i \in C$}
    \FOR{each $j \in L_i$}
        \STATE vertex $v = (i,j)$
        \STATE Add vertex $v$ to $G$
    \ENDFOR
\ENDFOR
\FOR{each pair of vertices $(i, j)$ and $(k,l)$ in $G$}
    \IF{i == k}
        \STATE Continue
    \ENDIF
    \STATE overlap $\leftarrow$ overlap between $j$ and $l$ calculated using Algorithm~\ref{alg:check_overlap}
    \IF{overlap $\geq b$}
        \STATE edge $e = ((i, j),(k,l))$
        \STATE weight of edge $w(e) = q_{ij}.A_i + q_{kl}.A_k$
        \STATE Add weighted edge $\{e,w(e)\}$ to $G$ 
    \ENDIF
\ENDFOR
\STATE max\_weight $\leftarrow$ $\max \{w(e)$ for edge $e \in G \}$
\FOR{each edge $e \in G$}
    \STATE $w(e) = \text{max\_weight} - w(e)$
\ENDFOR
\RETURN $G$
\end{algorithmic}
\end{algorithm}

\begin{lemma}
\label{thm:graph_construction_with_weights}
    Algorithm~\ref{alg:graph_construction_with_weights} constructs the compatibility graph for  given set $C$ of $N$ quantum circuits and the lists of their respective layouts in $\mathcal{O}((N.M)^2 \times n)$ where $M$ denotes the overall number of layouts, and $n$ is the length of the largest layout, or, in other words, the number of qubits in the largest circuit.
\end{lemma}

\emph{Proof}: See Appendix~\ref{app:graph_construction_with_weights}.

Scheduling the optimal number of circuits simultaneously, thus, boils down to finding the maximal clique from this graph. Since each vertex in a clique is connected to every other vertex, each of these circuits can be executed simultaneously. Each vertex layout is provided a weight which is the product of the mapomatic score for that layout and the normalized circuit area of the circuit. Every edge is associated with a weight which is the sum of the two weights of the associated vertices. However, recall that the lower the mapomatic score, the better (less noisy) is the layout. To convert this to a maximization problem, we find the largest edge weight, and subtract each edge weight from it. Thus, the edges corresponding to higher mapomatic score, i.e., more noisy layouts, now have lower weights, and vice versa. Hence, selecting the maximal clique from this graph  ensures selection of edges corresponding to lower mapomatic scores.

\begin{figure}[h!]
    \centering
    \includegraphics[scale=0.16]{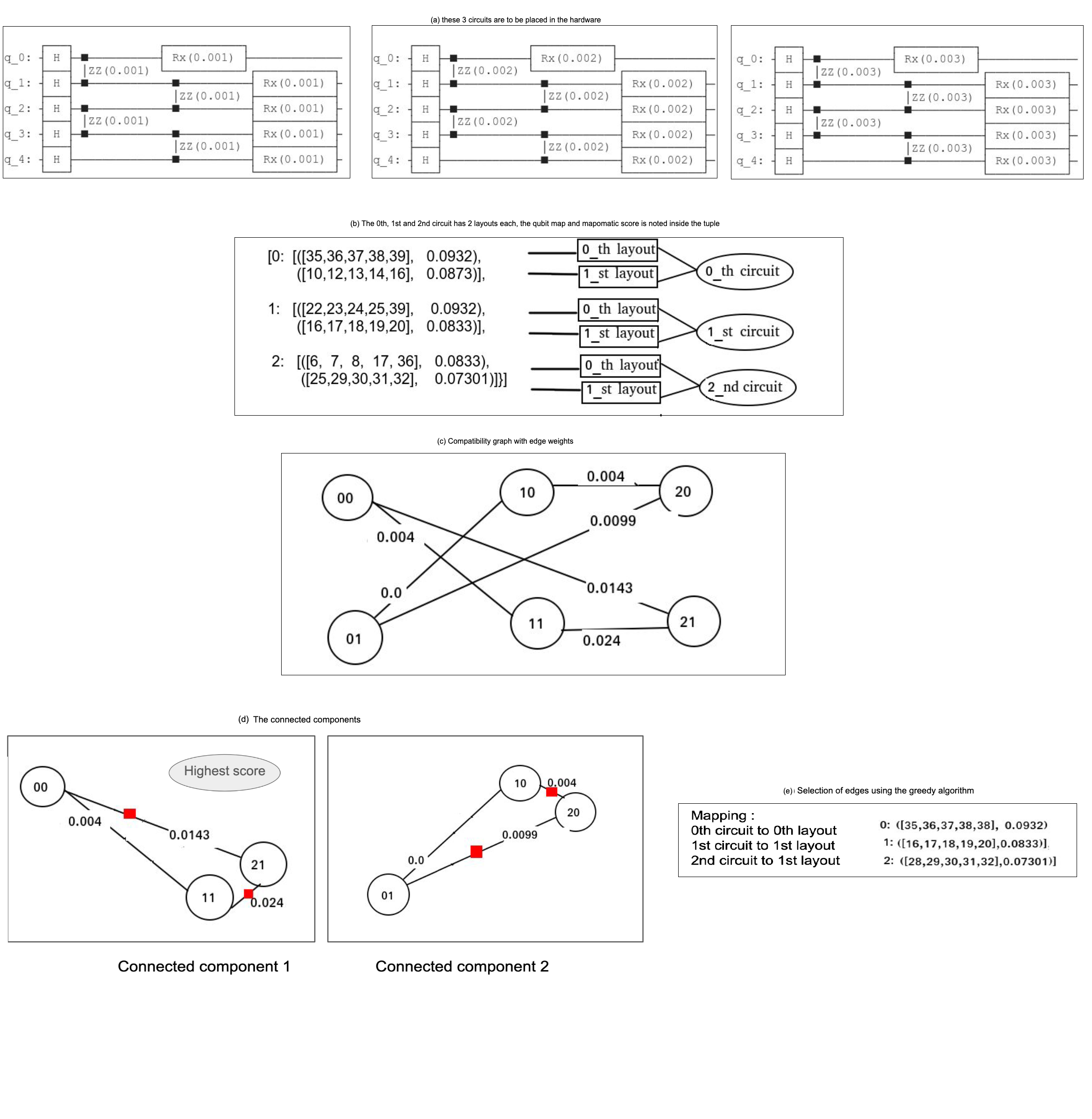}
    \caption{The entire workflow of our heuristic algorithm: (a) schematic  diagrams of three circuits that are to be placed in the hardware; (b) two possible layouts  for each of the three circuits and their corresponding mapomatic scores; (c) the compatibility graph with edge weights,  (d) the connected components of the graph and (e) greedy selection of edges in each of the components.}
    \label{fig:greedy}
\end{figure}

In Fig.~\ref{fig:greedy} (a) we take three example circuits, and show the construction of the compatibility graph in (b) and (c) of the same figure. We have selected the circuits to be of the same number of qubits and depth for this example, making $A_i = 1$ for all of them. In particular, Fig.~\ref{fig:greedy} (b) shows the scenario where each circuit has two compatible isomorphic layouts, and their mapomatic score. Usually, there will be many more such layouts for each circuit, but for this example we stick to two layouts for brevity. The compatibility graph is created from this information where each vertex $i,j$ corresponds to the layout $j$ for vertex $i$. From Algorithm~\ref{alg:graph_construction_with_weights}, no two vertices corresponding to the same vertex will have associated edge, since the same circuit is not to be placed twice. Therefore, no edge exists between any vertex with the same circuit index.

Two vertices are connected by an edge only if the layouts are not $b$-overlapping. For this example figure, we have selected $b = 1$. Thus, there is an edge between, say vertices $(00)$ and $(11)$, but not between $(00)$ and $(10)$ since they have a common qubit (39) in their layouts. The mapomatic scores for, say vertices $(00)$ and $(11)$, are $0.0932$ and $0.0833$. So, initially we assign the weight of the edge to be the sum of these two mapomatic scores, i.e., $0.1765$. After assigning weights to every edge similarly, we see the largest weight $0.1805$ is associated with the edge corresponding to vertices $(01)$ and $(10)$. Therefore, we subtract all the edge weights from this value, thus yielding the final weight of the edge corresponding to vertices $(00)$ and $(11)$ to be $0.004$. The other edges and their weights are similarly calculated.

A maximum clique in this compatibility graph provides the optimal noise-aware intra device scheduling. However, finding a maximum clique in any arbitrary graph is NP-Hard as well. In the following subsection we propose a greedy approach to find a \emph{maximal} clique in the compatibility graph for our scheduling problem.

\subsection{Greedy algorithm to find a \emph{maximal} clique in the compatibility graph}

\begin{algorithm}[h!]
\scriptsize
\caption{Finding maximal clique in a compatibility graph}
\label{alg:selection_process}

\begin{algorithmic}[1]

\REQUIRE compatibility graph $G = (V,E)$ where $E$ is the weighted edge list
\ENSURE maximal clique in $G$

\STATE $G_C \leftarrow$ the set of all connected components of $G$
\STATE max\_clique $\leftarrow \phi$
\STATE max\_clique\_weight $= 0$
\FOR{each $g \in G_C$}
    \STATE $V_g \subseteq V \leftarrow$ set of vertices in $g$, $E_g \subseteq E \leftarrow$ set of edges in $g$
    \STATE selected\_circuits $\leftarrow \phi$, selected\_layouts $\leftarrow \phi$, selected\_edges $\leftarrow \phi$
    \STATE $E_{g_{sorted}} \leftarrow$ sorted $E_g$ in the descending order of edge weight

    \FOR{each $e = (u,v) \in E_{g_{sorted}}$}
        \STATE $l_u, l_v \leftarrow$ layouts associated with $u$ and $v$ respectively
        \STATE $c_u, c_v \leftarrow$ circuits associated with $u$ and $v$ respectively
        
        \IF{selected\_edges is $\phi$}
            \STATE add $e$ to selected\_edges, $c_u$ and $c_v$ to selected\_circuits $l_u$ and $l_v$ to selected\_layouts, $e$ to selected\_edges
            
        \ELSIF{$c_u \in$ selected\_circuits \AND $c_v \in$ selected\_circuits}
            \STATE Continue

        \ELSIF{$c_u \notin$ selected\_circuits \AND $c_v \notin$ selected\_circuits}
            \STATE is\_connected $=$ True
            \FOR{all $c \in$ selected\_circuits}
              
                \IF{$(u,c) \notin E_g$ \OR $(v,c) \notin E_g$}
                    \STATE is\_connected $=$ False
                    \STATE \textbf{break}
                \ENDIF
            \ENDFOR
            \IF{is\_connected}
                \STATE add $e$ to selected\_edges, $c_u$ and $c_v$ to selected\_circuits $l_u$ and $l_v$ to selected\_layouts, $e$ to selected\_edges
            \ENDIF
            
        \ELSIF{$c_u \notin$ selected\_circuits \AND $l_u \in$ selected\_layouts}
            \STATE is\_connected $=$ True
            \FOR{all $c \in$ selected\_circuits}
                \IF{$(u,c) \notin E_g$}
                    \STATE is\_connected $=$ False
                    \STATE \textbf{break}
                \ENDIF
            \ENDFOR
            \IF{is\_connected}
                \STATE add $e$ to selected\_edges, $c_u$ and $c_v$ to selected\_circuits $l_u$ and $l_v$ to selected\_layouts, $e$ to selected\_edges
            \ENDIF
    
        \ELSIF{$c_v \notin$ selected\_circuits \AND $l_v \in$ selected\_layouts}
            \STATE is\_connected $=$ True
            \FOR{all $c \in$ selected\_circuits}
        
                \IF{$(v,c) \notin E_g$}
                    \STATE is\_connected $=$ False
                    \STATE \textbf{break}
                \ENDIF
            \ENDFOR
            \IF{is\_connected}
                \STATE add $e$ to selected\_edges, $c_u$ and $c_v$ to selected\_circuits $l_u$ and $l_v$ to selected\_layouts, $e$ to selected\_edges
            \ENDIF
        \ENDIF
      
    \ENDFOR
    \STATE clique $\leftarrow$ construct clique from selected\_edges and selected\_circuits
    \STATE weight $\leftarrow \sum_{e \in selected\_edges} w(e)$
    \IF{weight $>$ max\_clique\_weight}
        \STATE max\_clique\_weight $=$ weights, max\_clique $=$ clique
    \ENDIF
\ENDFOR
\RETURN max\_clique, max\_clique\_weight
\end{algorithmic}
\end{algorithm}

Algorithm~\ref{alg:selection_process} first determines the connected components of the compatibility graph. For each connected component it finds a maximal clique using a greedy method. It first selects the edge with the largest weight, and then keeps adding edges, sorted in descending order of weight, as long as the vertices are connected to all the vertices already selected (i.e., it is a clique). This ensures that the layouts selected are compatible with \emph{all} other layouts, and all the selected circuits can be scheduled simultaneously on the hardware. Note that this method generates one maximal clique for each connected component. Finally, the maximal clique with the largest weight among the ones found for each of the connected components is selected as the solution to the noise-aware intra-device scheduling.

\begin{lemma}
\label{thm:selection_process}
    Algorithm~\ref{alg:selection_process} finds a \emph{maximal} clique in the compatibility graph in $\mathcal{O}(|V| \cdot |E| + |E| \log |E| )$ 
\end{lemma}

\emph{Proof}: See Appendix~\ref{app:selection_process}.

In the next section, we present experimental results of our method to show the improvement in throughput obtained and the quality of the outcome.


\section{Experimental results}
\label{sec:results}

For our experiments with our proposed greedy method, we have considered 4 benchmarks circuits, namely Real Amplitude, Trotterized Clustered Unitary, QAOA, and Ripple carry adder. Although our formulation (Sections~\ref{sec:framework} and ~\ref{sec:heuristic}) do  not impose any constraints on the type of circuits that can be scheduled together,   we report here for only the circuits from the same family to study intra-device scheduling. A more rigorous experiment, with circuits from different families scheduled together, will be reported in another article separately.

For each circuit, we created its mirrored version. For a given circuit with unitary $U$,  a mirrored circuit of it can be obtained by appending $U^{\dagger}$ to the original circuit. This simple modification implies that the ideal outcome of the circuit is always $\ket{0}^{\otimes n}$, $n$ being the number of qubits in the circuit. The advantage of such a circuit is that the ideal outcome is known without any simulation. However, the disadvantage is that the depth of such a circuit is twice that of the original circuit, and is hence more amenable to noise. In Qiskit \cite{Qiskit}, it is necessary to put a barrier between $U$ and $U^{\dagger}$ in order to avoid simplification of the circuit to identity.

We first provide the rationale behind selecting the $\epsilon$ for layouts in our experiment (refer to Sec.~\ref{sec:framework}), and then show the fidelity obtained and the increase in throughput for a 27-qubit fake backend, and a 127-qubit IBM Quantum device.

\subsection{Selection of $\epsilon$ for the layouts}
It is expected that there is overlap between the layouts returned by  mapomatic. We have taken only the layouts having a score which is at least 50\% of the highest score and further  we check  for overlap among them and other conditions. If this percentage is increased, we can accommodate more circuits with less fidelity. 

  Table 1 gives a comparison of the values of fidelity for taking the best score, the worst score and the last of top $50\%$ in the noisy simulator of IBMQ Kolkata for 5-qubit circuits.

\begin{table*}[htb]
\label{tab:choosinglayout}
\centering
\caption{Comparison of values of fidelity for the best score, the worst score and the last of the top $50\%$ in noisy simulator of IBMQ Kolkata for 5-qubit circuits }
\begin{tabular}{|c|c|c|c|c|}
\hline
Benchmark &    2Q Depth  &  \multicolumn{3}{|c|}{Fidelity} \\
\cline{3-5}
 Circuit  & &   Best & Worst & Last of the top 50\%\\
\hline
Real Amplitude & 8& 0.944	&0.805	&0.925 \\
\hline
QAOA & 8 &0.879	&0.638	&0.842\\
\hline 
Trotterized & 18& 0.749	&0.29	&0.638\\

\hline 

\end{tabular}

\end{table*}

\subsection{Fidelity and hardware utilization in intra-device scheduling}

In Table 2, we consider the Real Amplitude circuits of different qubits to be run into the noisy simulator or IBMQ Kolkata (27-qubit) with and without using the intra-device scheduling. A total of 7 circuits of each type was chosen for the experiments. Here 2 circuits are placed in the  hardware simultaneously exhibiting a better throughput and resource utilization. We show that the values of fidelity where we are using intra device scheduling is almost reaching the fidelity if the circuits are one to one mapped in the best available hardware. In our experiments we have used buffer $b$=1, i.e., between two circuit mapped there should be a gap of at least 1 qubit. This is to minimize the cross talk where it  is maximum if two circuits are placed without a single qubit barrier \cite{van2023probabilistic}.

\begin{table}[htb]
\label{tab:fakekol}
\centering
\caption{Fidelity for different sized Real Amplitude circuits with and without using our intra-device scheduling  to be run on Noisy IBMQ simulator with the noise profile and coupling map of 27-qubit IBMQ Kolkata  }
\begin{tabular}{|c|c|c|c|}
\hline
Circuit size &  Circuit Count & $Fidelity_{Int}$ & $Fidelity_{NoInt}$ \\
\# qubits & & & \\

\hline
3&   & 0.9542837452 & 0.959822345 \\
5&  7 & 0.9244571429 & 0.959822345 \\
7& & 0.8468928571 & 0.862323176 \\
10&  & 0.6612723723 & 0.675571234 \\

\hline 

\end{tabular}

\end{table}

\begin{figure}[htb]
    \centering
    \includegraphics[scale=0.2]{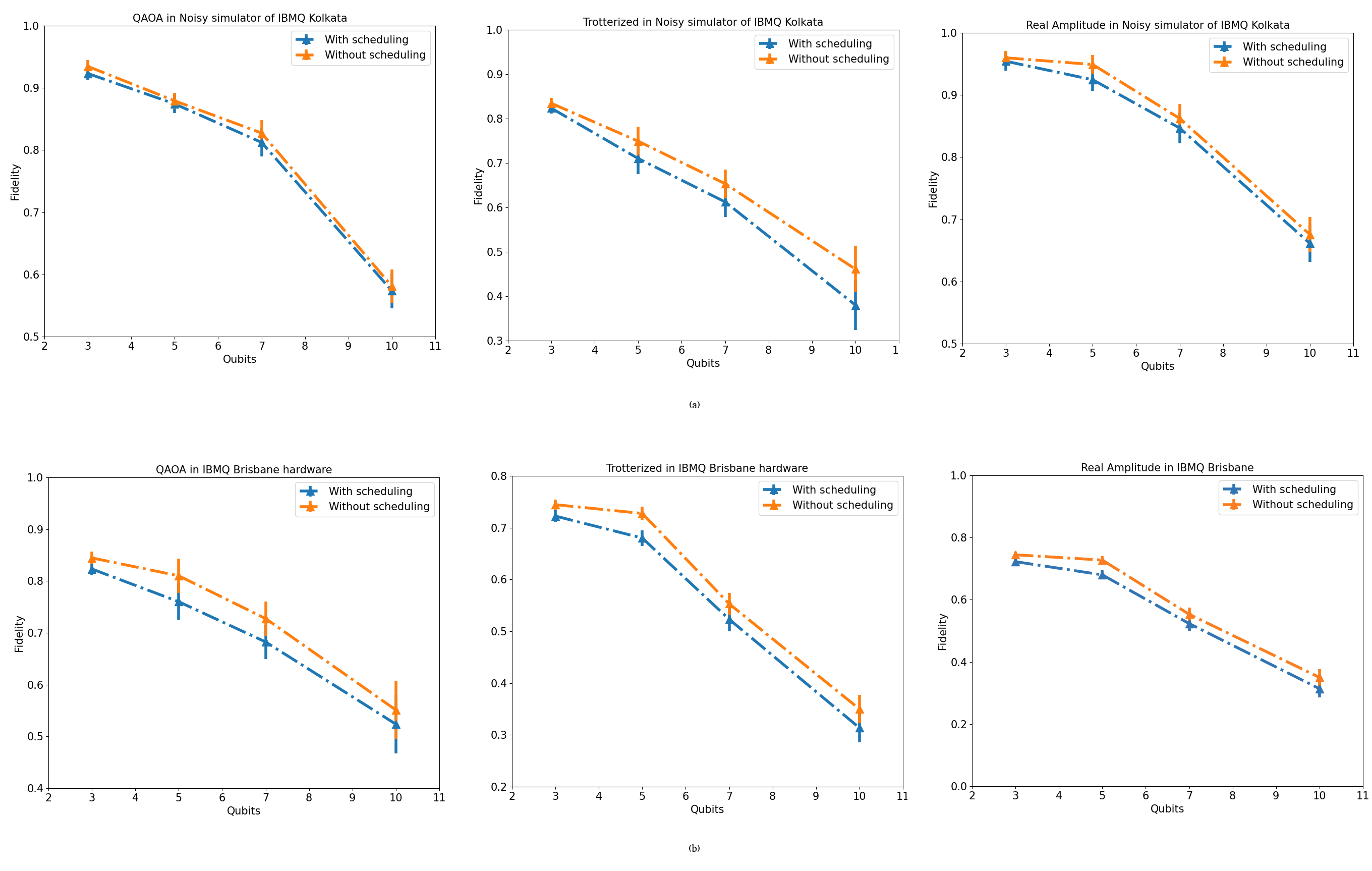}
    \caption{Fidelity (along with the mean and standard deviation) for benchmark circuits (QAOA, Trotterized, Real Amplitude) with and without using our intra-device scheduling  executed in (a) Noisy IBMQ simulator with the noise profile and coupling map of 27-qubit IBMQ KOlkata,  (b) 127-qubit IBMQ Brisbane hardware. }
    \label{fig:plot_Result}
\end{figure}

In Fig. \ref{fig:plot_Result}, we consider the   benchmark circuits QAOA, Trotterized, Real Amplitude having different number of qubits to be run on (a) Noisy IBMQ simulator with the noise profile and coupling map of 27-qubit IBMQ Kolkata , and (b) 127-qubit IBMQ Brisbane hardware. With intra-device scheduling, 3 circuits are placed in the  hardware to be executed simultaneously and thereby exhibiting a better throughput and resource utilization. We show that the values of fidelity where we are using intra-device scheduling is almost reaching the fidelity if the circuits are  mapped one by one in the best available hardware. We have also given the mean and standard deviation for each of the points, where Mean is the average value of the dataset, indicating the central point and   Standard Deviation is the measure of the spread or dispersion of the dataset around the mean.

\begin{table*}[htb]
\label{hwutil}
\centering
\caption{Hardware utilization in intra device scheduling }
\begin{tabular}{|c|c|c|c|}
\hline
Circuit size &  Hardware size $m$ & \# circuits placed simultaneously & Gain w.r.t. time \\

\hline

7& 27 & 2 & 2x \\
10& 27 & 2 & 2x\\

7& 127 & 3& 3x \\
10& 127 & 3 & 3x\\

\hline 

\end{tabular}

\end{table*}

If we have included more circuits in the hardware simultaneously of course the hardware utilization would be better but we constrained our solution with the top 50\% score of the best score from the hardware layout. Note thatL: Number of swap gates for each of these layout will be equal because the mapomatic solution uses graph isomorphism to cimpute the possible layouts. The time and utilizatation can be improved with the expense of fidelity.

\section{Conclusion}
\label{sec:conclusion}

In this paper, we addressed the critical challenge of optimizing quantum circuit scheduling to enhance the throughput and efficiency of quantum computing hardware. By drawing analogies to the classical bin packing problem, we demonstrated the NP-Hard nature of our problem, which involves placing multiple quantum circuits onto quantum processing units while considering the inherent noise and limited qubit connectivity. Our proposed solution, using integer linear programming or the greedy heuristic based solution on compatibility graphs and maximal cliques, effectively balances the trade-off between noise reduction and throughput optimization. The experimental results showed significant improvements in time utilization, achieving 2x and 3x better efficiency for 27-qubit and 127-qubit hardware, respectively. These findings highlight the potential of intra-device scheduling to maximize the performance of NISQ-era quantum computers, paving the way for more reliable and scalable quantum computing solutions in the future.

It is intuitive that if we increase the number of layouts allowed for further processing from 50\% of top scores, then the utilization will be better where as the fidelity can be worse. This trade-off between number of layout vs fidelity will be studied experimentally as a future work. Moreover how the buffer distance affects the fidelity is also a work which is left for future studies.

\bibliographystyle{unsrt}
\bibliography{main}

\appendix

\section{Proof of Theorem~\ref{thm:overlap}}
\label{app:overlap}
\begin{theorem}
    The overlap between the two layouts $l_1$ and $l_2$ can be determined by calculating the distance between the boundary qubits of these two layouts only.
\end{theorem}

By definition, if $b \in l$ is a boundary qubit in layout $l$, then $\exists$ $q \notin l$ such that $q \in neighbour(b)$, and the overlap between two layouts is the minimum distance between any two qubits from the two layouts. Let $B_l \subseteq l$ be the set of all boundary qubits in layout $l$. Therefore, a shortest path between some $q_l \in l$, but $\notin B_l$ and $q \notin l$ must contain some $b \in B_l$. Hence, $d(b,q) < d(q_l,q)$. Therefore, the overlap between two layouts can be determined by calculating the distance between the boundary qubits of these two layouts only.

\section{Proof of Lemma~\ref{thm:find_boundary}}
\label{app:find_boundary}

Algorithm ~\ref{alg:find_boundary} iterates through each qubit $q \in l$ exactly once to determine whether $q \in B_l$ where $B_l$ denotes the set of boundary qubits of layout $l$. For this, the algorithm checks whether $n_q \in l$ $\forall$ $n_q \in neighbour(q)$. The time required for this is $\mathcal{O}(d_q)$ where $d_q$ denotes the degree of $q$. Majority of the current quantum devices conform to a planar graph architecture. Therefore, the degree of the qubits are bounded. Since $d_q$ does not depend on the length of the layout, the overall time requirement to determine the boundary qubits of $l$ is \(\mathcal{O}(|l|)\).

\section{Proof of Lemma~\ref{thm:check_overlap}}
\label{app:check_overlap}

The algorithm \ref{alg:check_overlap} first checks for $0$-overlap between $l_i$ and $l_j$. This overlap can be determined in $\mathcal{O}(\min \{|l_i|, |l_j|\})$. When the two qubits are $0$-overlapping, the algorithm determines the boundary qubits for both the layouts in $\mathcal{O}(\max \{|l_i|, |l_j|\})$ if done in parallel. Let $B_i \subseteq l_i$ and $B_j \subseteq l_j$ denote the set of boundary qubits of $l_i$ and $l_j$ respectively, where $|B_i| = k_i$ and $|B_j| = k_j$. Now, the algorithm checks for the distance between every $b_i \in B_i$ and $b_j \in B_j$ to determine the buffer $b$ in $\mathcal{O}(k_i \times k_j)$.
    
Thus, the overall time complexity of the algorithm is $\mathcal{O}(\max\{|l_i|, |l_j|\} + k_i \times k_j)$, where $\max\{|l_i|, |l_j|\}$ accounts for the time complexity of finding the boundaries of $l_i$ and $l_j$, followed by the time $k_i \times k_j$ required for the pairwise distance checking between boundary qubits of the two layouts.

\section{Proof of Lemma~\ref{thm:graph_construction_with_weights}}
\label{app:graph_construction_with_weights}

In order to construct the compatibility graph, the algorithm first generates the set of vertices, each of which is a (circuit, layout) pair, in $\mathcal{O}(N.M)$ by iterating through each layout for each circuit. Next, for each pair of vertices, it calculates overlap between the two associated layouts $l_1$ and $l_j$ in time $\mathcal{O}(\max \{|l_i|, |l_j|\} + k_i \times k_j)$ as per Algorithm~\ref{alg:check_overlap} (see Lemma~\ref{thm:overlap} for details of the notation) to generate the compatible edges. The time to calculate overlap is dominated by the length of the largest layout, say $n$. Since, there are $\mathcal{O}(\begin{pmatrix}
    N.M \\
    2
\end{pmatrix})$ possible pairs of vertices, the overall time requirement for generating the set of edges is $\mathcal{O}(\begin{pmatrix}
    N.M \\
    2
\end{pmatrix} \times n)$. Finally, subtracting the weight of each edge from the maximum weight requires $\mathcal{O}(\begin{pmatrix}
    N.M \\
    2
\end{pmatrix})$. Hence, the overall time required to generate the compatibility graph is $\mathcal{O}(N.M) + \mathcal{O}(\begin{pmatrix}
    N.M \\
    2
\end{pmatrix} \times n) + \mathcal{O}(\begin{pmatrix}
    N.M \\
    2
\end{pmatrix}) = \mathcal{O}((N.M)^2 \times n)$.

\section{Two necessary lemmata and their proofs}

\begin{lemma}
\label{app:lemma_a}
If $M = \sum_g M_g $, where $M_g \geq 0$ $\forall$ $g$, then  $M \log M \geq \sum_g M_g \log M_g$
\end{lemma}

\emph{Proof}:
\begin{eqnarray*}
    M &=& \sum_g M_g \\
    M \log M &=& \sum_g M_g \log \sum_g M_g \\
    &=& M_1 \log \sum_g M_g + M_2 \sum_g M_g +... \\
    &\geq&  M_1 \log M_1  + M_2 \log  M_2 + ....
\end{eqnarray*}

where the final inequality follows since $M_g \geq 0$ $\forall$ $g$ and logarithm is a monotonically increasing function.

\begin{lemma}
\label{app:lemma_b}
If $M = \sum_g M_g $ and $N = \sum_g N_g $, where $M_g \geq 0$ and $N_g \geq 0$ $\forall$ $g$, then $N \cdot M \leq \sum_g N_g \cdot M_g$.
\end{lemma}

\emph{Proof}:
\begin{eqnarray*}
    N \cdot M &=& (\sum_g M_g) \cdot (\sum_g N_g) \\
    &=& \sum_g M_g \cdot N_g + \sum_{g \neq h} M_g \cdot N_h \\
    &\leq& \sum_g M_g \cdot N_g
\end{eqnarray*}
where the final inequality follows since $M_g \geq 0$ and $N_g \geq 0$ $\forall$ $g$.

\section{Proof of Theorem~\ref{thm:selection_process}}
\label{app:selection_process}

Let $G = (V, E)$ be the compatibility graph. First, the algorithm identifies the connected components of the graph. This can be achieved using a Breadth-First-Search in $\mathcal{O}(|V| + |E|)$ time. Let $g = (V_g, E_g)$ denote a connected component. For each connected component, the algorithm first sorts the edges in $\mathcal{O}(|E_g| \log |E_g|)$ time. Next, for each edge, the algorithm checks whether the two associated vertices are connected to all the vertices already present in the constructed clique. For each edge, this can be performed in $\mathcal{O}(V_g)$. Therefore, performing this check for all the edges requires $\mathcal{O}(|V_g| \cdot |E_g|)$ time.

This exercise is repeated for all the connected components. Therefore, the overall time requires is $\sum_g \mathcal{O}(|V_g| \cdot |E_g|)$ + $\mathcal{O}(|E_g| \log |E_g|)$. Now from Lemma~\ref{app:lemma_a} and Lemma~\ref{app:lemma_b}: $\sum_g \mathcal{O}(|V_g| \cdot |E_g|)$ + $\mathcal{O}(|E_g| \log |E_g|) $ = $\mathcal{O}(|V| \cdot |E|)$ + $\mathcal{O}(|E| \log |E|)$. Finally, the weight of the clique for each connected component can be calculated in $\mathcal{O}(|E_g|)$, thus requiring a total of $\sum_g \mathcal{O}(|E_g|) = \mathcal{O}(|E|)$ for all the components. 
  
Hence the time complexity of Algorithm~\ref{alg:selection_process} is 
  \begin{center}
      $\mathcal{O}(|V| + |E|)$ + $\mathcal{O}(|V| \cdot |E|)$ + $\mathcal{O}(|E| \log |E|)$ + $\mathcal{O}(|E|)  = \mathcal{O}(|V| \cdot |E| + |E| \log |E| )$ 
  \end{center}

\end{document}